# Enhancement of metallicity by Na doping in La$_3$Ni$_2$O$_{7+\delta}$

Yingying Gao[1,2], W. Zhou[2,*], W. H. Guo[1,2], Chunqiang Xu[4], H. F. Chen[1,2], Z. D. Han[2], Xiaofeng Xu[3,*], Yinzhong Wu[1,*] and B. Qian[2,*]

1. *School of Physical Science and Technology, Suzhou University of Science and Technology, Suzhou 215009, China*

2. *School of Electronic and Information Engineering, Suzhou University of Technology, Changshu, 215500, China*

3. *School of Physics, Zhejiang University of Technology, Hangzhou 310023, China*

4. *School of Physical Science and Technology, Ningbo University, Ningbo 315211, China*

**The observation of high-$T_c$ superconductivity in bilayer nickelate La$_3$Ni$_2$O$_7$ under high pressure provides a new venue for exploring novel unconventional superconductors and elucidating the mechanism of high-$T_c$ superconductivity. Subsequently, numerous chemical substitution studies have been reported, aiming to stabilize superconductivity at ambient pressure, or significantly reduce the pressure threshold required for its occurrence. Here, we report the comprehensive study on sodium (Na) doping in the Ruddlesden-Popper nickelate La$_3$Ni$_2$O$_{7+\delta}$, where Na$^+$ substitutes for La$^{3+}$ at the A-site with varying doping concentrations. The structural, thermal, magnetic, and electronic transport properties of as-synthesized polycrystalline samples were systematically investigated. X-ray diffraction (XRD) analysis reveals that Na doping induces a structural transition from the '327' *Amam* phase to the '4310' *Bmab* phase when $x \geq 0.075$, which is further corroborated by thermogravimetric analysis (TGA) measurements. Substitution of La$^{3+}$ with Na$^+$ gives rise to a gradual expansion of the '327' phase lattice. Meanwhile, resistivity measurements indicate that the density wave (DW) transition is marginally suppressed and metallicity is significantly enhanced. Upon the application of pressure, DW transition can be further suppressed, whereas the low-$T$ insulating behaviors remain insensitive to pressure. These results offer critical insights into the roles of elemental substitution and charge carrier doping in steering the competing electronic phases in layered nickelates.**

**Corresponding Email:** wei.zhou@szut.edu.cn; xuxiaofeng@zjut.edu.cn; yzwu@usts.edu.cn; njqb@szut.edu.cn

## 1. Introduction

The discovery of high-temperature (high-$T_c$) superconductivity (HTS) in bilayer nickelate $La_3Ni_2O_7$ under high pressure, with a critical temperature ($T_c$) approaching 80 K, has opened a new frontier in the search for unconventional superconductivity beyond the paradigms of cuprate and iron-based superconductors [1-6]. This material class belongs to the $n=2$ Ruddlesden-Popper (RP) phase ($La_{n+1}Ni_nO_{3n+1}$), featuring a unique crystal structure composed of double $NiO_2$ layers separated by rock-salt La-O blocks. Its electronic structure exhibits a multi-orbital character, where both Ni $3d x^2-y^2$ and $3d z^2$ orbitals contribute significantly to the Fermi surface [7-11], in stark contrast to the single-band physics of infinite-layer nickelates or cuprates [12, 13]. This complexity, coupled with its remarkably high $T_c$, suggests a distinct superconducting mechanism likely involving strong interlayer coupling and Hund's metal physics [8-11, 14-16].

A central theme emerging from recent studies is the profound sensitivity of superconductivity in $La_3Ni_2O_{7+\delta}$ to its chemical composition and structural details [14, 17]. The oxygen non-stoichiometry ($\delta$) of the parent compound has been identified to be a critical tuning parameter; a comprehensive phase diagram reveals that superconductivity emerges for $\delta \geq 0.00$ at pressures above ~25 GPa, adjacent to an insulating state at highly negative $\delta$ and an Anderson localized regime at intermediate oxygen deficiency [17]. Furthermore, the presence and concentration of oxygen vacancies, particularly at the inner apical sites bridging the two $NiO_6$ octahedra along the *c*-axis, are found to be detrimental to the superconducting state, as they disrupt the crucial interlayer magnetic exchange and orbital hybridization [14, 18]. Concurrently, chemical substitution at the La site has proven to be a powerful strategy to modulate its properties [4, 5, 19-24]. Doping with smaller rare-earth ions (e.g., Sm or Nd) can induce lattice compression, which stabilizes the high-$T_c$ phase and even leads to superconductivity signatures at lower pressures, while leaving the primary density wave (DW) transition largely unaltered [6, 19, 20, 23]. These findings underscore that both charge carrier density and lattice strain are key variables governing the delicate balance between competing electronic orders, such as spin/charge density waves [25, 26] and superconductivity, in this system.

Alkali metal doping, such as sodium ($Na^+$) doping, presents a promising yet underexplored avenue for further investigation. Substitution of $La^{3+}$ by $Na^+$ constitutes an aliovalent process that acts primarily as hole doping, directly altering the Ni valence state and overall band filling without introducing apparent magnetic moments. This doping strategy differs markedly from the doping effect associated with isovalent rare-earth substitution (e.g., $Sm^{3+}$ for $La^{3+}$). Na doping thus provides a clean experimental approach to investigate the role of hole concentration in the phase diagram of $La_3Ni_2O_7$, potentially driving the system away from the commensurate 7.5 $3d$ electrons per Ni ion [1, 6, 17]. Such a systematic study would serve as a crucial complement to existing investigations on oxygen content and rare-earth doping, enabling a more comprehensive understanding of how different types of chemical pressure and carrier doping influence the stability of the superconductivity and competing DW states.

In this work, we present a systematic study of polycrystalline $La_{3-x}Na_xNi_2O_{7+\delta}$ samples, focusing on the evolution of their crystal structure, electronic transport, and magnetic properties as a function of Na content ($x$) at ambient pressure. Our primary objective is to elucidate how hole doping via Na substitution modifies the ground state of $La_3Ni_2O_7$ and to evaluate its potential for stabilizing or enhancing superconductivity under high pressure. This study thereby contributes to the broader effort of mapping the intricate electronic phase space of this emerging high-$T_c$ nickelate family.

## 2. Experiments

All polycrystalline $La_{3-x}Na_xNi_2O_{7+\delta}$ samples with various nominal doping levels were synthesized via the sol-gel method as reported previously [6, 23]. Stoichiometric amounts of raw materials, i.e., $La(NO_3)_3 \cdot 6H_2O$ (99.9% Aladdin), $NaNO_3$ (99.9% Aladdin), and $Ni(NO_3)_2 \cdot 6H_2O$ (99.9% Aladdin), were mixed and dissolved in deionized water with an appropriate amount of citric acid. The resulting solution was heated on a hot plate and stirred at 180°C for 4 hours to form a homogeneous green gel. The gel was subsequently transferred to a drying oven and baked at 180°C overnight to yield a green powder. This green powder was slowly heated in air to 800°C and held for 12 hours to remove organic components. Finally, the obtained black powder was grounded, pressed into pellets, and further heated slowly to 1100°C and maintained at this temperature for 36 hours to obtain the final polycrystalline samples.

Powder X-ray diffraction (XRD) data were collected at room temperature using a diffractometer with Cu $K\alpha$ radiation. The chemical composition and microstructure were characterized by energy-dispersive X-ray spectroscopy (EDS) and transmission electron microscopy (TEM), respectively (Supplementary materials, SI). Thermogravimetric analysis (TGA) was performed under a 5% $H_2$/Ar gas flow at 50 mL/min from the room temperature to 900 °C, with a heating rate of 5 °C/min. Magnetic susceptibility data were measured using a Quantum Design (QD) superconducting quantum interference device (SQUID) magnetometer. For all electrical resistivity measurements, a standard four-probe method was applied and the data were acquired on a physical properties measurement system (PPMS). High-pressure electrical resistance measurements were carried out on PPMS using a piston cylinder–type pressure cell, using liquid Daphene 7373 as the pressure transmitting medium. The actual applied pressure was calibrated by measuring the pressure-dependent $T_c$ of a standard Pb reference sample.

## 3. Results and discussion

Figure 1 shows the XRD data of the as-synthesized polycrystalline samples. For $x=0$, most of the major peaks can be indexed with the theoretical XRD pattern calculated based on the reported '327' *Amam* phase (RP phase $La_{n+1}Ni_nO_{3n+1}$, $n=2$) (Fig.1 (f)) [1, 6, 23, 27]. With increasing $x$, the XRD peaks generally shift toward lower diffraction angles, indicating the expansion of the crystal lattice. However, when $x$ reaches 0.075, several new peaks emerge as marked by the red diamonds in Figs. 1 (a) and (f). Meanwhile, the peaks at ~23.5 and ~32 degrees exhibit splitting behavior (Figs. 1 (b) and (c)). By comparing with the reported structures for RP nickelates [27-31], these new peaks are suggested to originate from the orthorhombic '4310' *Bmab* phase (RP phase $La_{n+1}Ni_nO_{3n+1}$, $n=3$) (Fig. 1(f)). Based on our analysis, the peaks indicated by the red arrows in Figs. 1 (b) and (c) belong to the '327' *Amam* phase, and the strongest peak around 32.3 degree for $x \geq 0.075$ is attributed to the '4310' *Bmab* phase. Accordingly, the dominant phases of the as-synthesized polycrystalline samples change to the '4310' *Bmab* phase when $x \geq 0.075$. This change of the dominant phase is further supported by the subsequent TGA measurements. The XRD patterns for $0.075 \leq x \leq 0.5$ are shown in the SI. For $x \geq 0.2$, no further peak shifts are observed with the increase of $x$, implying a possible saturation of Na doping. Figs. 1 (d) and (e) show the lattice parameters and unit cell volume calculated from the XRD patterns. For $x=0$, the lattice parameters $a$ and $b$ are slightly larger than the reported values, which is presumably due to the incorporation of additional oxygen, as indicated by the later TGA data. With increasing $x$, the unit cell tends to expand, which is unexpected given that the ionic radius of $Na^+$ is smaller than that of $La^{3+}$. Nevertheless, a similar lattice expansion has also been observed in Na-doped $La_2CuO_4$, which is explained by the larger effective ionic radius of nine-coordinated Na vs. La [32].

The TGA curves presented in Fig. 2(a) illustrate the weight loss behavior of as-synthesized polycrystalline samples ($x$ = 0, 0.025, 0.05, and 0.1) as a function of temperature up to 800 °C. All samples undergo a gradual initial weight loss starting at approximately 350 °C, followed by a more pronounced mass reduction between 500 °C and 650 °C. This two-stage weight loss behavior is consistent with previous reports [6, 33], which implies the samples have decomposed into $La_2O_3$, Ni, and Na at the plateaus of the TGA curves in the high-temperature region. Based on the weight loss, the calculated oxygen contents $7+\delta$ for $x$ = 0, 0.025, and 0.05 are listed in Table 1. It can be seen that the oxygen content decreases notably upon Na doping. The oxygen content obtained for $x=0$ is considerably higher than the previous reported value of ~6.93 [6, 33], which may induce lattice expansion due to inclusion of additional oxygen atoms, consistent with the larger lattice parameters obtained from XRD measurements. A larger weight loss is observed for the sample $x$ = 0.1. It is worth noting that the theoretical weight loss of $La_4Ni_3O_{10}$ is 7.18%, which is larger than the corresponding value of 6.19% for $La_3Ni_2O_7$. Accordingly, combined with the above XRD analysis, the observed larger weight loss for $x$ = 0.1 also supports the formation of the '4310' phase. The temperature-dependent magnetic susceptibility ($\chi$) curves of as-synthesized polycrystalline samples, measured from 2 K to 300 K under a magnetic field of 1 T, are presented in Fig. 2 (b). All samples exhibit a broad minimum in $\chi$ around 80–100 K, followed by a gradual increase with increasing temperature. A similar feature has been reported previously for undoped $La_3Ni_2O_{7+\delta}$ [6, 23, 27]. Na doping does not significantly alter the magnetic properties of $La_3Ni_2O_{7+\delta}$.

Table 1 Oxygen content of the as-synthesized $La_{3-x}Na_xNi_2O_{7+\delta}$ samples ($x$=0, 0.025, 0.05) calculated based on TGA. $M_{loss}$ values are obtained from Fig. 2 (a).

| $x$ | 0 | 0.025 | 0.05 |
|---|---|---|---|
| $M_{loss}$ | 6.43% | 6.37% | 6.44% |
| $7+\delta$ | 7.10 | 7.06 | 7.08 |

Figure 3 (a) presents the ambient-pressure resistivity curves for the Na-doped '327' samples ($0 \leq x \leq 0.05$). For the undoped sample ($x = 0$), the resistivity exhibits a broad minimum around 142 K, followed by a gradual increase upon further cooling, indicative of a crossover from metallic to insulating behavior at low temperatures. The resistivity anomaly at ~142 K has been previously attributed to a DW transition [1, 6, 23]. With Na doping, the resistivity is drastically reduced, and the DW transition temperature $T_{DW}$ determined from the resistivity minimum near the transition is slightly suppressed. Fig. 3 (b) shows the normalized resistivity ($\rho/\rho_{300\,K}$) curves. As observed, compared to the $x = 0$ sample, $\rho/\rho_{300\,K}(T)$ curves of Na-doped samples become steeper above $T_{DW}$, while the insulating behavior below $T_{DW}$ is strongly weakened. This phenomenon indicates a remarkable enhancement of metallic character upon Na doping, which may originate from the increased carrier concentration induced by Na substitution. Fig. 3 (c) and (d) show the resistivity curves for $x \geq 0.075$ samples whose main phase transforms to the '4310' phase. In comparison with the '327' phase, the resistivity magnitude of the "4310" phase is lower. Notably, the $\rho(T)$ curves do not evolve monotonically with further Na doping, and $T_{DW}$ changes little with different $x$ values, which may be associated with the saturation of Na doping as evidenced from the XRD results. The doping dependence of $T_{DW}$, shown in the inset of Fig. 3 (d), is in good agreement with that of the residual resistivity ratio (RRR≡ $\rho_{2\,K}/\rho_{300\,K}$), suggesting an intimate relationship between the DW transition and the low-$T$ insulating behavior. The temperature derivative of resistivity

(d$\rho$/d$T$) further highlights the evolution of the DW transition with varying $x$ (Fig. 3 (f)).

Given the enhanced metallicity by Na doping, we further investigated the effect of pressure on Na-doped samples to examine whether superconductivity can be induced at low pressures. To maintain a stable hydrostatic pressure environment, a liquid pressure-transmitting medium was used. The corresponding results are shown in Fig. 4. For both the $x = 0.05$ and $x = 0.1$ samples, the resistance decreases monotonically, and the DW transition gradually moves toward lower temperatures with increasing pressure. Nevertheless, the low-$T$ insulating behavior is very robust against pressure. By fitting the low-$T$ insulating behavior using a thermally activated model, the energy gaps under different pressures were extracted and shown in the inset of Fig. 4 (a). As seen, the energy gap remains nearly unchanged under different pressures, suggesting that the low-$T$ insulating state is insensitive to pressure, in sharp contrast to the suppression of $T_{DW}$. Fig. 4 (b) shows the enlarged view of the normalized resistance ($R/R_{300 K}$) curves. With the suppression of $T_{DW}$, the onset temperature of the low-$T$ insulating behavior ($T_{insulating}$) gradually moves toward high temperatures as pressure increases. Eventually, $T_{DW}$ and $T_{insulating}$ appear to converge. A similar behavior is also observed in the $x = 0.1$ sample (Fig. 4 (c)). Fig. 4 (d) summarizes the pressure dependence of $T_{DW}$ and $T_{insulating}$ for two compositions. The pressure suppression rate of $T_{DW}$ is −21.8 K/GPa for $x = 0.05$ and −26.9 K/GPa for $x = 0.1$. For $x = 0.05$, the absolute value of $T_{DW}$ suppression rate is lower than that (~ -25.7 K/GPa) of pure La$_3$Ni$_2$O$_{7+\delta}$ [25], implying that while $T_{DW}$ is slightly suppressed by Na doping, the suppression of DW transition becomes more difficult. Future higher pressure experiments will be performed to explore whether pressure-induced superconductivity can be realized in La$_{3-x}$Na$_x$Ni$_2$O$_{7+\delta}$, particularly for the $x$=0.025 and $x$=0.05 compositions.

Figure 5 summarizes the effect of Na doping on polycrystalline La$_3$Ni$_2$O$_{7+\delta}$ in the phase diagram. Above the DW transition, all samples display metallic behavior. With increasing Na doping level $x$, the DW transition is gradually suppressed, the metallicity is significantly enhanced, and the low-$T$ insulating region shrinks. For $x \geq 0.075$, the dominant phase transforms from the '327' $Amam$ phase to the '4310' $Bmab$ phase. Notably, the metallicity of the '4310' phase below the DW transition is significantly enhanced compared to that of the '327' phase.

## 4. Conclusion

In conclusion, we have systematically investigated the effect of Na doping on polycrystalline $La_3Ni_2O_{7+\delta}$. We demonstrated that in $La_{3-x}Na_xNi_2O_{7+\delta}$, pure '327' phase can only be stabilized within a very narrow Na composition range. For $x \geq 0.075$, the main phase transforms to the orthorhombic '4310' *Bmab* phase. Nevertheless, Na substitution for La leads to a pronounced enhancement of metallicity and the suppression of the DW transition. Upon applying pressure, the DW transition can be further suppressed, whereas the low-$T$ insulating behavior shows little sensitivity to pressure. As a direction of future work, it would be highly desirable to explore whether high-temperature superconductivity can be realized under higher pressures in these Na-doped '327' polycrystalline samples with improved metallicity.

## Acknowledgments

This work was sponsored by the National Natural Science Foundation of China (Grant no. 11704047, U1832147, 12274369, 12304071). A portion of work was supported by Zhejiang Provincial Natural Science Foundation of China (Grant No. LZ25A040003).

## Reference

[1] H. Sun, M. Huo, X. Hu, J. Li, Z. Liu, Y. Han, L. Tang, Z. Mao, P. Yang, B. Wang, J. Cheng, D.-X. Yao, G.-M. Zhang and M. Wang, Signatures of superconductivity near 80 K in a nickelate under high pressure, Nature, **621**, 493-498 (2023).

[2] J. Hou, P.-T. Yang, Z.-Y. Liu, J.-Y. Li, P.-F. Shan, L. Ma, G. Wang, N.-N. Wang, H.-Z. Guo, J.-P. Sun, Y. Uwatoko, M. Wang, G.-M. Zhang, B.-S. Wang and J.-G. Cheng, Emergence of High-Temperature Superconducting Phase in Pressurized $La_3Ni_2O_7$ Crystals, Chinese Physics Letters, **40**, 117302 (2023).

[3] Y. Zhang, D. Su, Y. Huang, Z. Shan, H. Sun, M. Huo, K. Ye, J. Zhang, Z. Yang, Y. Xu, Y. Su, R. Li, M. Smidman, M. Wang, L. Jiao and H. Yuan, High-temperature superconductivity with zero resistance and strange-metal behaviour in $La_3Ni_2O_{7-\delta}$, Nature Physics, **20**, 1269-1273 (2024).

[4] M. Zhang, C. Pei, Q. Wang, Y. Zhao, C. Li, W. Cao, S. Zhu, J. Wu and Y. Qi, Effects of

Pressure and Doping on Ruddlesden-Popper phases $La_{n+1}Ni_nO_{3n+1}$, Journal of Materials Science & Technology, **185**, 147-154 (2024).

[5] N. Wang, G. Wang, X. Shen, J. Hou, J. Luo, X. Ma, H. Yang, L. Shi, J. Dou, J. Feng, J. Yang, Y. Shi, Z. Ren, H. Ma, P. Yang, Z. Liu, Y. Liu, H. Zhang, X. Dong, Y. Wang, K. Jiang, J. Hu, S. Nagasaki, K. Kitagawa, S. Calder, J. Yan, J. Sun, B. Wang, R. Zhou, Y. Uwatoko and J. Cheng, Bulk high-temperature superconductivity in pressurized tetragonal $La_2PrNi_2O_7$, Nature, **634**, 579-584 (2024).

[6] G. Wang, N. N. Wang, X. L. Shen, J. Hou, L. Ma, L. F. Shi, Z. A. Ren, Y. D. Gu, H. M. Ma, P. T. Yang, Z. Y. Liu, H. Z. Guo, J. P. Sun, G. M. Zhang, S. Calder, J. Q. Yan, B. S. Wang, Y. Uwatoko and J. G. Cheng, Pressure-Induced Superconductivity In Polycrystalline $La_3Ni_2O_{7-\delta}$, Physical Review X, **14**, 011040 (2024).

[7] J. Yang, H. Sun, X. Hu, Y. Xie, T. Miao, H. Luo, H. Chen, B. Liang, W. Zhu, G. Qu, C.-Q. Chen, M. Huo, Y. Huang, S. Zhang, F. Zhang, F. Yang, Z. Wang, Q. Peng, H. Mao, G. Liu, Z. Xu, T. Qian, D.-X. Yao, M. Wang, L. Zhao and X. J. Zhou, Orbital-dependent electron correlation in double-layer nickelate $La_3Ni_2O_7$, Nature Communications, **15**, 4373 (2024).

[8] F. Lechermann, J. Gondolf, S. Bötzel and I. M. Eremin, Electronic correlations and superconducting instability in $La_3Ni_2O_7$ under high pressure, Physical Review B, **108**, L201121 (2023).

[9] Z. Luo, X. Hu, M. Wang, W. Wú and D.-X. Yao, Bilayer Two-Orbital Model of $La_3Ni_2O_7$ under Pressure, Physical Review Letters, **131**, 126001 (2023).

[10] Y. Zhang, L.-F. Lin, A. Moreo and E. Dagotto, Electronic structure, dimer physics, orbital-selective behavior, and magnetic tendencies in the bilayer nickelate superconductor $La_3Ni_2O_7$ under pressure, Physical Review B, **108**, L180510 (2023).

[11] C. Lu, Z. Pan, F. Yang and C. Wu, Interlayer-Coupling-Driven High-Temperature Superconductivity in $La_3Ni_2O_7$ under Pressure, Physical Review Letters, **132**, 146002 (2024).

[12] D. Li, K. Lee, B. Y. Wang, M. Osada, S. Crossley, H. R. Lee, Y. Cui, Y. Hikita and H. Y. Hwang, Superconductivity in an infinite-layer nickelate, Nature, **572**, 624-627 (2019).

[13] P. Puphal, T. Schäfer, B. Keimer and M. Hepting, Superconductivity in infinite-layer and Ruddlesden–Popper nickelates, Nature Reviews Physics, **8**, 70-85 (2026).

[14] Y.-B. Liu, J.-W. Mei, F. Ye, W.-Q. Chen and F. Yang, s±-Wave Pairing and the

Destructive Role of Apical-Oxygen Deficiencies in $La_3Ni_2O_7$ under Pressure, Physical Review Letters, **131**, 236002 (2023).

[15] Y. Gu, C. Le, Z. Yang, X. Wu and J. Hu, Effective model and pairing tendency in the bilayer Ni-based superconductor $La_3Ni_2O_7$, Physical Review B, **111**, 174506 (2025).

[16] Z. Liu, M. Huo, J. Li, Q. Li, Y. Liu, Y. Dai, X. Zhou, J. Hao, Y. Lu, M. Wang and H.-H. Wen, Electronic correlations and partial gap in the bilayer nickelate $La_3Ni_2O_7$, Nature Communications, **15**, 7570 (2024).

[17] Y. Ueki, H. Sakurai, H. Nagata, K. Yamane, R. Matsumoto, K. Terashima, K. Hirose, H. Ohta, M. Kato and Y. Takano, Phase Diagram of Pressure-Induced High Temperature Superconductor $La_3Ni_2O_{7+\delta}$, Journal of the Physical Society of Japan, **94**, 013703 (2024).

[18] J. Li, D. Peng, P. Ma, H. Zhang, Z. Xing, X. Huang, C. Huang, M. Huo, D. Hu, Z. Dong, X. Chen, T. Xie, H. Dong, H. Sun, Q. Zeng, H.-k. Mao and M. Wang, Identification of superconductivity in bilayer nickelate $La_3Ni_2O_7$ under high pressure up to 100 GPa, National Science Review, **12**, nwaf220 (2025).

[19] Z. Pan, C. Lu, F. Yang and C. Wu, Effect of Rare-Earth Element Substitution in Superconducting $R_3Ni_2O_7$ under Pressure, Chinese Physics Letters, **41**, 087401 (2024).

[20] F. Li, Z. Xing, D. Peng, J. Dou, N. Guo, L. Ma, Y. Zhang, L. Wang, J. Luo, J. Yang, J. Zhang, T. Chang, Y.-S. Chen, W. Cai, J. Cheng, Y. Wang, Y. Liu, T. Luo, N. Hirao, T. Matsuoka, H. Kadobayashi, Z. Zeng, Q. Zheng, R. Zhou, Q. Zeng, X. Tao and J. Zhang, Bulk superconductivity up to 96 K in pressurized nickelate single crystals, Nature, **649**, 871-878 (2026).

[21] M. Xu, S. Huyan, H. Wang, S. L. Bud'ko, X. Chen, X. Ke, J. F. Mitchell, P. C. Canfield, J. Li and W. Xie, Pressure-Dependent "Insulator–Metal–Insulator" Behavior in Sr-Doped $La_3Ni_2O_7$, Advanced Electronic Materials, **10**, 2400078 (2024).

[22] B. Hao, M. Wang, W. Sun, Y. Yang, Z. Mao, S. Yan, H. Sun, H. Zhang, L. Han, Z. Gu, J. Zhou, D. Ji and Y. Nie, Superconductivity in Sr-doped $La_3Ni_2O_7$ thin films, Nat Mater, **24**, 1756-1762 (2025).

[23] J.-J. Feng, T. Han, J.-P. Song, M.-S. Long, X.-Y. Hou, C.-J. Zhang, Q.-G. Mu and L. Shan, Unaltered density wave transition and pressure-induced signature of superconductivity in Nd-doped $La_3Ni_2O_7$, Physical Review B, **110**, L100507 (2024).

[24] Q. Zhong, J. Chen, Z. Qiu, J. Li, X. Huang, P. Ma, M. Huo, H. Dong, H. Sun and M. Wang, Evolution of the superconductivity in pressurized $La_{3-x}Sm_xNi_2O_7$, arXiv:2510.13342v1, (2025).

[25] R. Khasanov, T. J. Hicken, D. J. Gawryluk, V. Sazgari, I. Plokhikh, L. P. Sorel, M. Bartkowiak, S. Bötzel, F. Lechermann, I. M. Eremin, H. Luetkens and Z. Guguchia, Pressure-enhanced splitting of density wave transitions in $La_3Ni_2O_{7-\delta}$, Nature Physics, **21**, 430-436 (2025).

[26] D. Zhao, Y. Zhou, M. Huo, Y. Wang, L. Nie, Y. Yang, J. Ying, M. Wang, T. Wu and X. Chen, Pressure-enhanced spin-density-wave transition in double-layer nickelate $La_3Ni_2O_{7-\delta}$, Science Bulletin, **70**, 1239-1245 (2025).

[27] C. D. Ling, D. N. Argyriou, G. Wu and J. J. Neumeier, Neutron Diffraction Study of La3Ni2O7: Structural Relationships Among n=1, 2, and 3 Phases $La_{n+1}Ni_nO_{3n+1}$, Journal of Solid State Chemistry, **152**, 517-525 (2000).

[28] J. Zhang, H. Zheng, Y.-S. Chen, Y. Ren, M. Yonemura, A. Huq and J. F. Mitchell, High oxygen pressure floating zone growth and crystal structure of the metallic nickelates $R_4Ni_3O_{10}$ (R=La, Pr), Physical Review Materials, **4**, 083402 (2020).

[29] H. Sakakibara, M. Ochi, H. Nagata, Y. Ueki, H. Sakurai, R. Matsumoto, K. Terashima, K. Hirose, H. Ohta, M. Kato, Y. Takano and K. Kuroki, Theoretical analysis on the possibility of superconductivity in the trilayer Ruddlesden-Popper nickelate $La_4Ni_3O_{10}$ under pressure and its experimental examination: Comparison with La3Ni2O7, Physical Review B, **109**, 144511 (2024).

[30] M. Shi, Y. Li, Y. Wang, D. Peng, S. Yang, H. Li, K. Fan, K. Jiang, J. He, Q. Zeng, D. Song, B. Ge, Z. Xiang, Z. Wang, J. Ying, T. Wu and X. Chen, Absence of superconductivity and density-wave transition in ambient-pressure tetragonal $La_4Ni_3O_{10}$, Nature Communications, **16**, 2887 (2025).

[31] H. Li, X. Zhou, T. Nummy, J. Zhang, V. Pardo, W. E. Pickett, J. F. Mitchell and D. S. Dessau, Fermiology and electron dynamics of trilayer nickelate $La_4Ni_3O_{10}$, Nature Communications, **8**, 704 (2017).

[32] C. C. Torardi, M. A. Subramanian, J. Gopalakrishnan and A. W. Sleight, Alkali-metal substituted $La_2CuO_4$: Structures of $La_{2-x}M_xCuO_4$ (M Na, K; x~0.2), Physica

C-Superconductivity and Its Applications, **158**, 465 (1989).

[33] Q. Liu, J. Yi, Y. Gu, Y. Shi, M. Zhou, J. Yuan, H. He, L. Chen, G. Chen, J. Yu and Z. Ren, Emergence of tetragonal phase in oxygen-annealed Co-doped $La_3Ni_2O_{7+delta}$, Journal of Solid State Chemistry, **351**, 125528 (2025).

# Figures

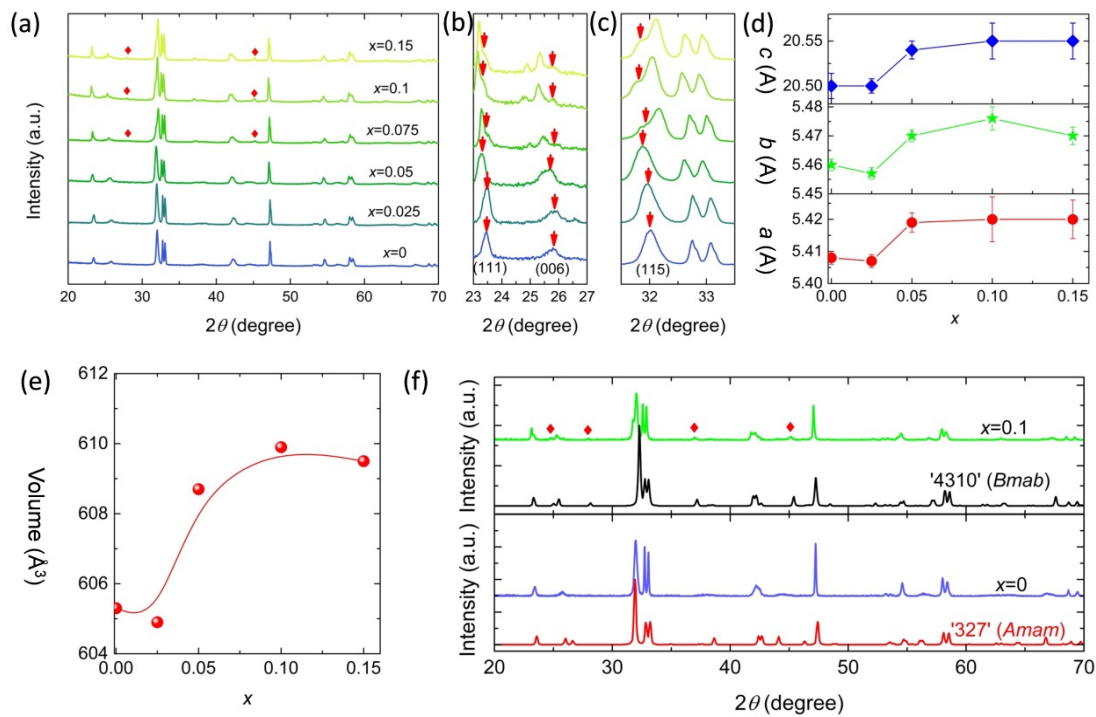

**Figure 1** (a) XRD patterns for the as-synthesized $La_{3-x}Na_xNi_2O_7$ polycrystalline samples. For $x \geq 0.075$, the main phase has changed to the orthorhombic '4310' *Bmab* phase. The red diamonds indicate the peaks only belonging to the '4310' *Bmab* phase. (b) and (c) The enlarged views of the peaks at specific angle ranges. For $x \geq 0.075$, the (111), (006), and (115) peaks for the '327' *Amam* phase all show apparent splitting. The peaks indicated by the red arrows are attributed to the original '327' *Amam* phase. (d) The calculated lattice parameters $a$, $b$, and $c$ for the '327' *Amam* phase versus the doping level $x$. (e) Evolution of unit cell volume for the '327' *Amam* phase with $x$. (f) Comparison between the experimental XRD patterns of samples $x=0$ and $x=0.1$ and the theoretical XRD patterns reported for the '327' *Amam* phase and the '4310' *Bmab* phase. The red diamonds represent the peaks that only belong to the '4310' *Bmab* phase.

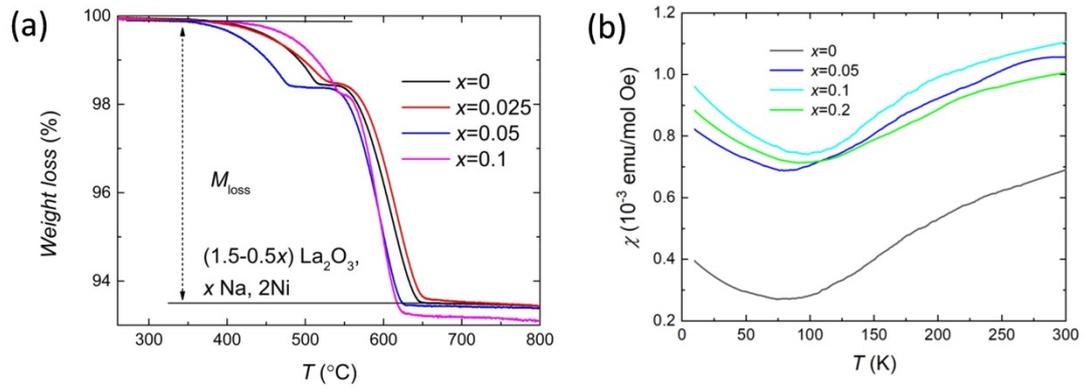

**Figure 2** (a) Thermogravimetric analysis (TGA) curves of the as-synthesized polycrystalline samples. $M_{loss}$ is defined as the difference between the initial and the final plateaus on the TGA curves. (b) Temperature dependence of susceptibility ($\chi$) measured under a magnetic field of 1 T.

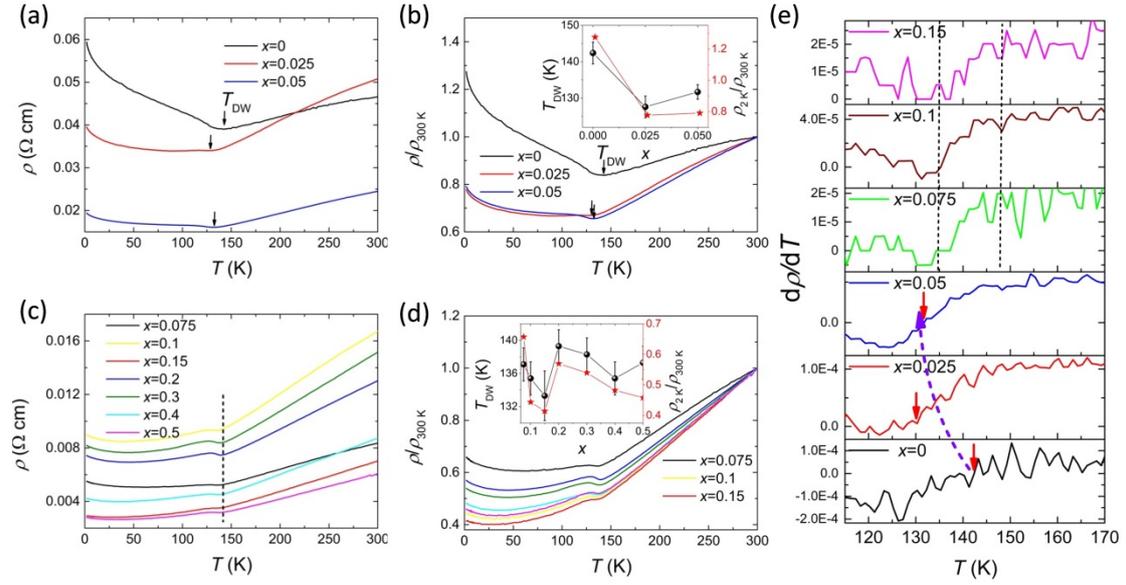

**Figure 3** (a) Temperature dependence of resistivity ($\rho(T)$ curves) for $x \leq 0.05$. $T_{DW}$ is defined as the DW transition temperature at the resistivity minimum. (b) $\rho(T)$ curves normalized by the room-temperature resistivity $\rho_{300\,K}$. Inset: Doping dependence of $T_{DW}$ and the residual resistivity ratio (RRR=$\rho_{2\,K}/\rho_{300\,K}$). (c) and (d) $\rho(T)$ curves and normalized resistivity ($\rho/\rho_{300\,K}(T)$) curves for $0.075 \leq x \leq 0.5$. Inset in (d): $T_{DW}$ and $\rho_{2\,K}/\rho_{300\,K}$ as a function of $x$ for $0.075 \leq x \leq 0.5$. (e) The derivative of resistivity with respect to temperature ($d\rho/dT$) near the DW transition. The red arrows denote the positions for $d\rho/dT=0$ for samples $x$=0, 0.025, and 0.05, where the resistivity locates at the minimum.

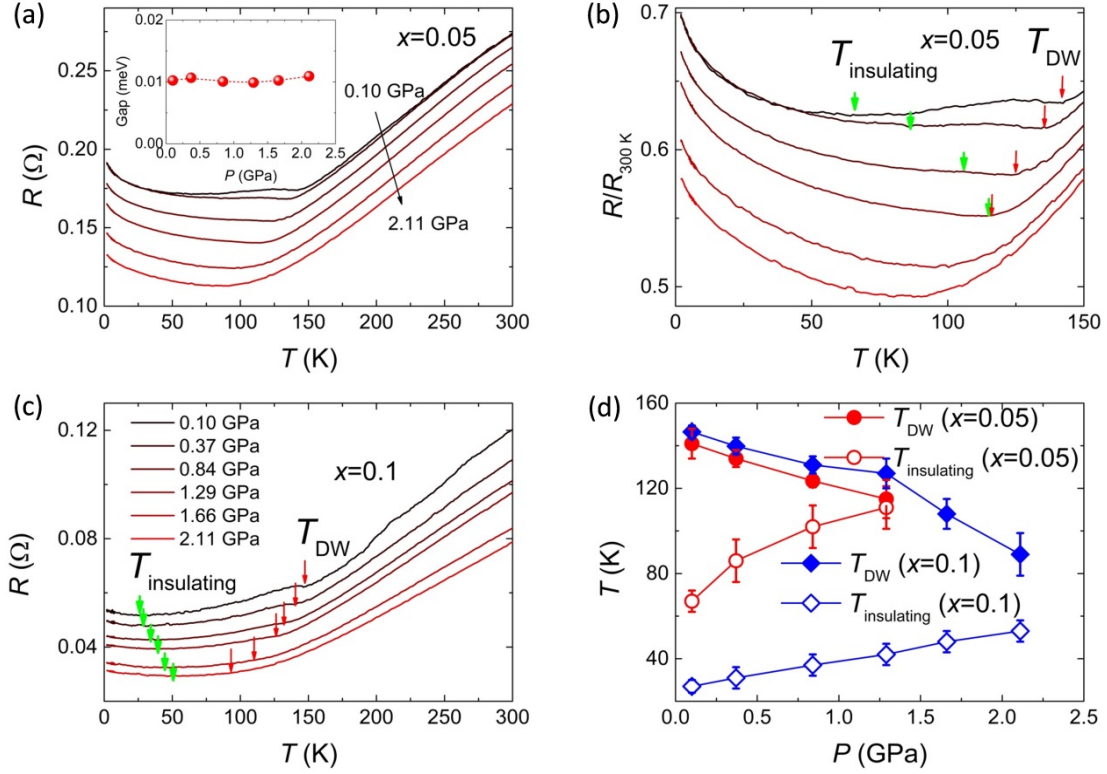

**Figure 4** (a) Temperature dependence of resistance for sample $x = 0.05$ under different pressures. Inset: Pressure dependence of the energy gap ($\Delta$) obtained from the thermally activated model $R = R_0 \exp(\frac{\Delta}{k_B T})$ for the low-$T$ (below 10 K) resistance. (b) The blow-up view of the resistance curves below 150 K. The resistance is normalized by the value at 300 K. $T_{DW}$ represents the transition temperature of the density wave transition. $T_{insulating}$ defines the onset temperature of low-$T$ insulating behavior. (c) Temperature dependence of resistance for the sample $x = 0.1$ under different pressures. (d) Pressure dependence of $T_{DW}$ and $T_{insulating}$ for both $x = 0.05$ and $x = 0.1$.

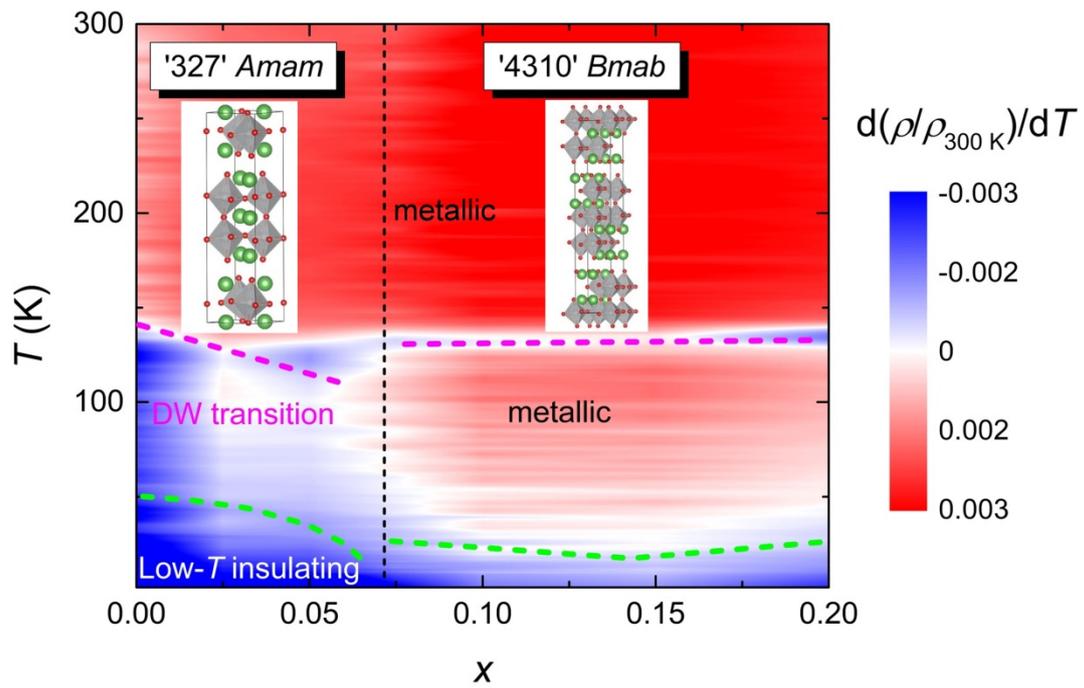

Figure 5 Doping dependent phase diagram of the as-synthesized La$_{3-x}$Na$_x$Ni$_2$O$_7$ polycrystalline samples. Insets show the atomic crystal structures of the '327' *Amam* phase and the '4310' *Bmab* phase